%% file: ckm2012_behari.tex
\documentclass[12pt]{article}
\usepackage{epsfig}
\usepackage{amsmath}

\input econfmacros.tex
\def\Title#1{\begin{center} {\Large {\bf #1} } \end{center}}
\renewcommand\footnotemark{\ }

\begin{document}

\Title{CDF results on $b \rightarrow s \mu \mu$ decays}

\bigskip\bigskip


\begin{raggedright}  
{\it Satyajit Behari\index{Behari, S.}\\
Fermi National Accelerator Laboratory\\
Batavia, IL 60510, USA} \\
\bigskip
{\footnotesize Proceedings of CKM 2012,
the 7th International Workshop on the CKM Unitarity Triangle, \\
University of Cincinnati, USA, 28 September - 2 October 2012} \\
\bigskip\bigskip
\end{raggedright}

\section{Introduction}
\label{sec:intro}
Rare decays of bottom hadrons mediated by the flavor-changing neutral current 
(FCNC) process $b \to s \mu^+ \mu^-$ occur in the standard model (SM) through 
higher order (loop) amplitudes. A variety of beyond-the-standard-model (BSM) 
theories, on the other hand, favor enhanced rates for these FCNC decays, where 
heavy exotic particles may participate in the loops. These processes are thus
very interesting tools to search for BSM physics. In particular, these 
three-body decays provide observables sensitive to NP, {\em e.g.} the branching
ratios, their dependence on the di-muon mass distribution and the angular
distributions of the decay products.

We summarize recent $b \to s \mu^+ \mu^-$ results from the CDF experiment 
based on the full 9.6 fb$^{-1}$ dataset collected in $p\bar{p}$ collisions at
$\sqrt(s)$ = 1.96 TeV. The decays analyzed are; $B^+ \to K^+ \mu^+ \mu^-$,
$B^0 \to K^{*0}(892) \mu^+ \mu^-$, $B^0 \to K^0_s \mu^+ \mu^-$, 
$B^+ \to K^{*+}(892) \mu^+ \mu^-$, $B^0_s \to \phi \mu^+ \mu^-$, and
$\Lambda^0_b \to \Lambda \mu^+ \mu^-$. The latter two decays were first
observed by CDF~\cite{CdfPhiLam} in 2011. From an angular analysis of the
$B \to K^* \mu^+ \mu^-$ decays we also present updated results on the 
transverse polarization and T-odd CP asymmetries reported 
earlier~\cite{CdfPolAcp}.

\section{Branching Ratios}
\label{sec:BR}
\begin{figure}[htb]
\begin{center}
\epsfig{file=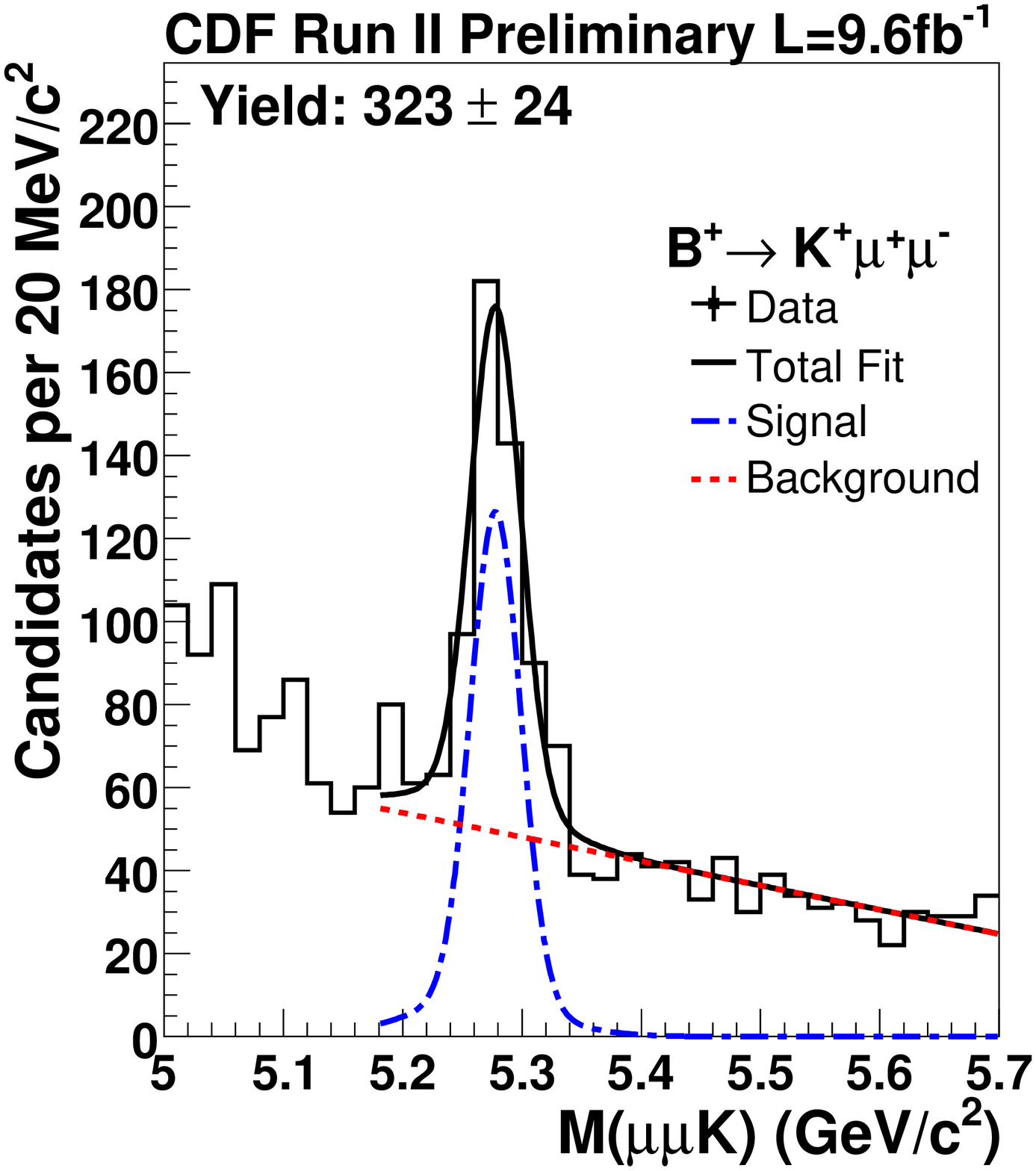,width=0.27\textwidth}
\epsfig{file=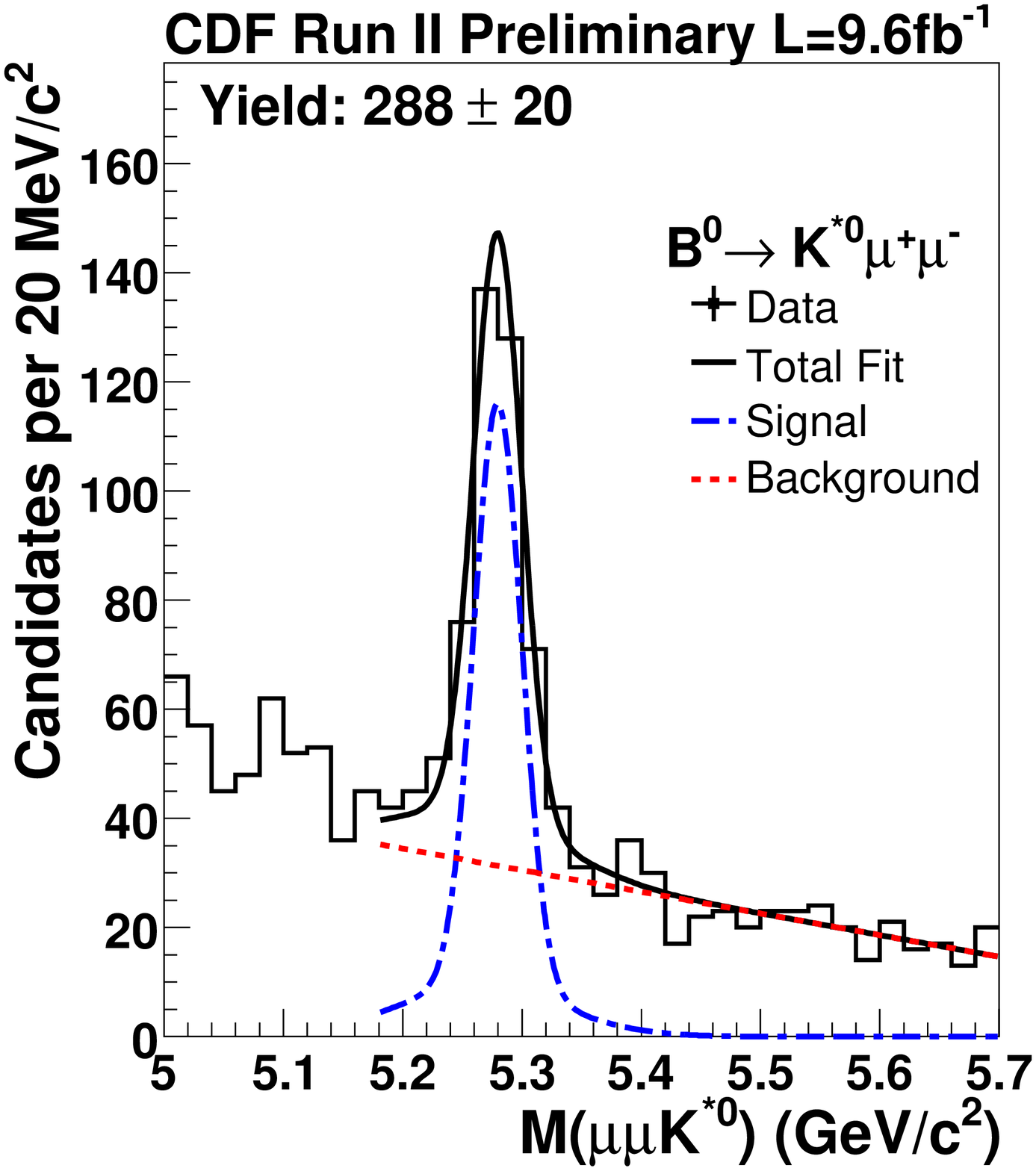,width=0.27\textwidth}
\epsfig{file=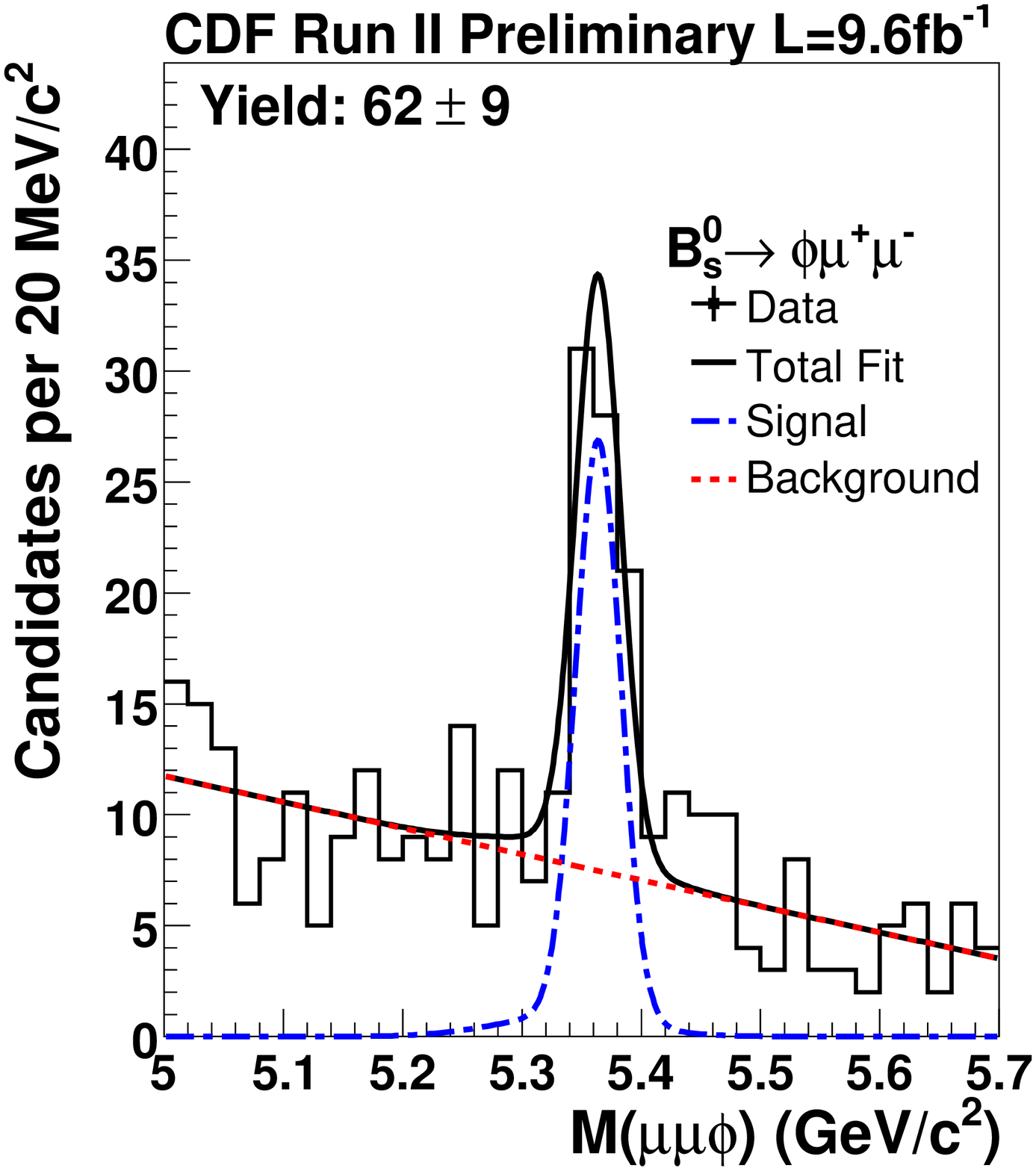,width=0.27\textwidth}
\epsfig{file=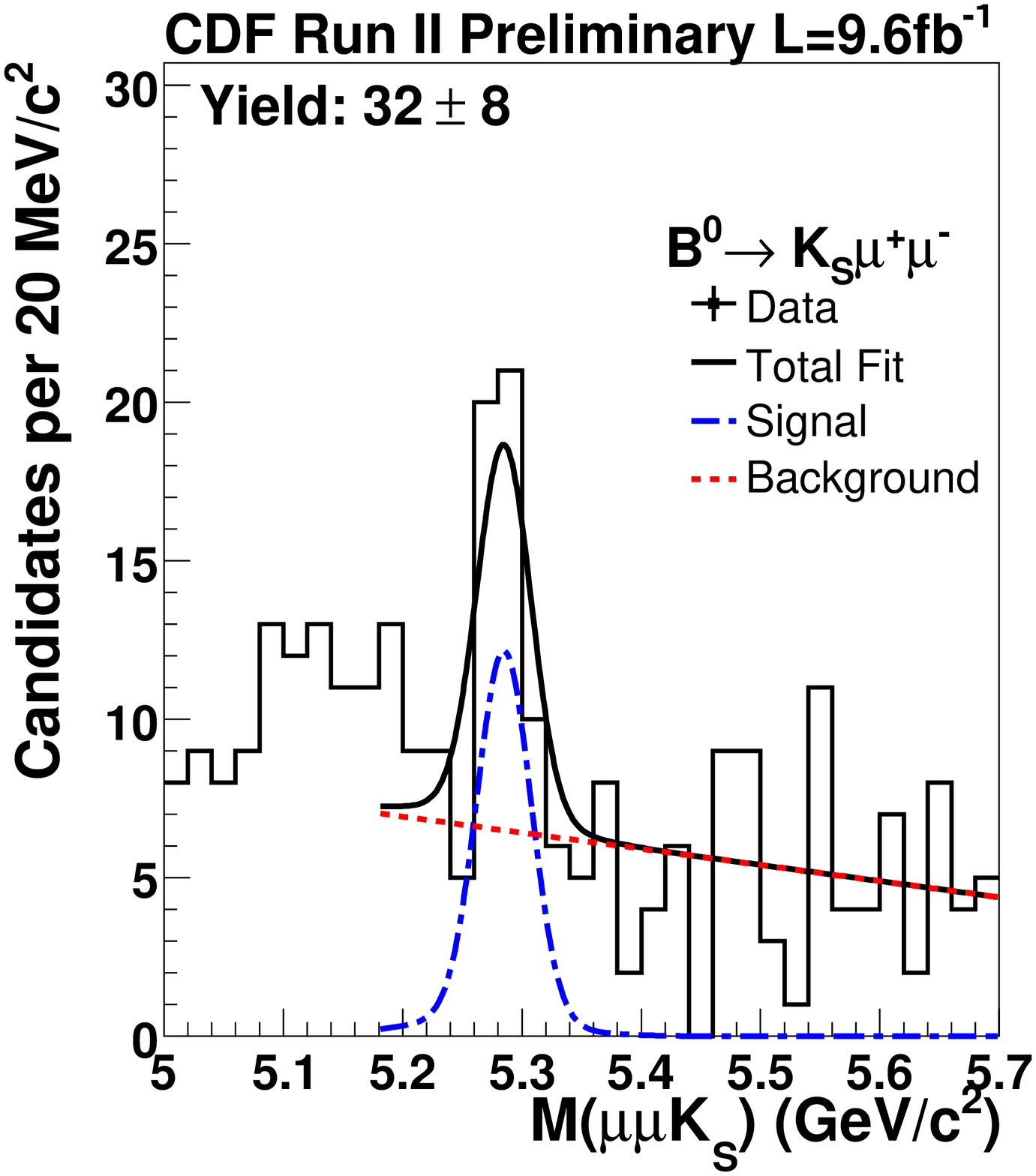,width=0.27\textwidth}
\epsfig{file=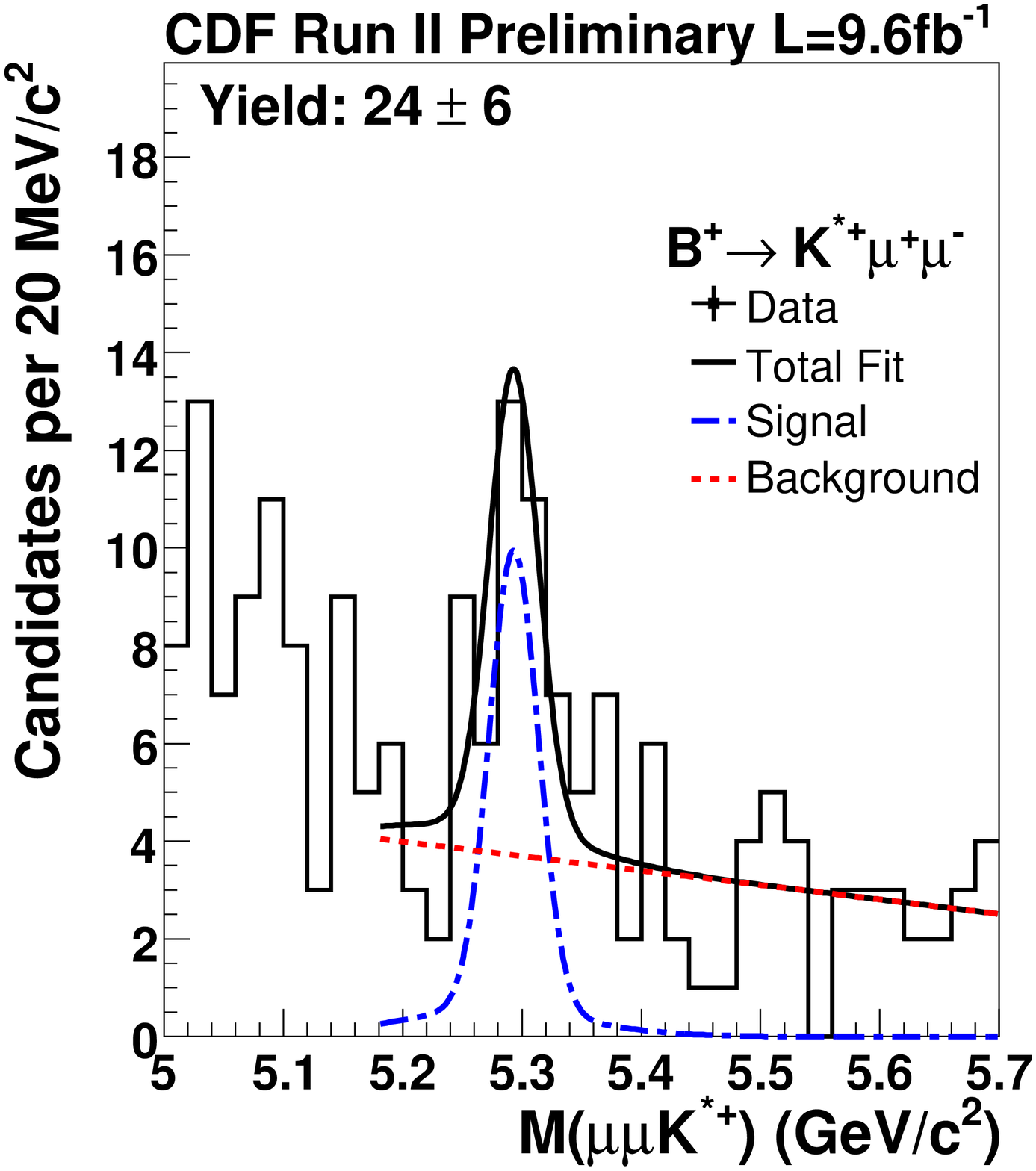,width=0.27\textwidth}
\epsfig{file=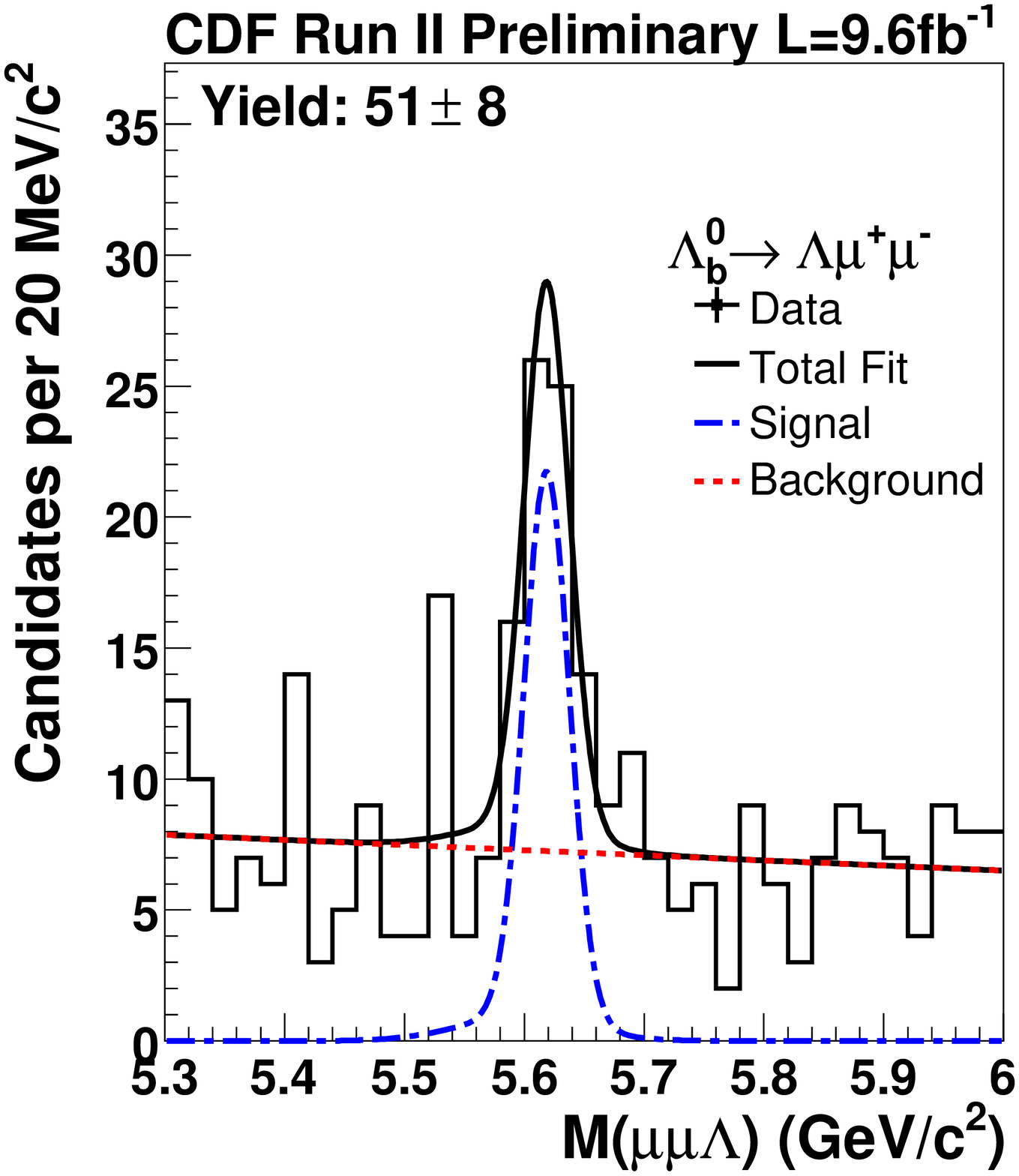,width=0.27\textwidth}
\caption{Signal yields in $B^+ \to K^+ \mu^+ \mu^-$,
$B^0 \to K^{*0}(892) \mu^+ \mu^-$, $B^0 \to K^0_s \mu^+ \mu^-$, 
$B^+ \to K^{*+}(892) \mu^+ \mu^-$, $B^0_s \to \phi \mu^+ \mu^-$, and
$\Lambda^0_b \to \Lambda \mu^+ \mu^-$ modes.}
\label{fig:yields}
\end{center}
\end{figure}

The signal yields of the analyzed rare decays are obtained by unbinned 
maximum log-likelihood fits to the invariant mass distributions, shown 
in Figure~\ref{fig:yields}.

The measured relative branching ratios with resepect the corresponding
reference channels are:
\begin{eqnarray}
\nonumber
\mathcal{B}(B^+ \to K^+ \mu^+ \mu^-)/\mathcal{B}(B^+ \to J/\psi K^+)
  = [0.44 \pm 0.03(\mathrm{stat}) \pm 0.02(\mathrm{syst})] \times 10^{-3}, \\
\nonumber
\mathcal{B}(B^0 \to K^{*0} \mu^+ \mu^-)/\mathcal{B}(B^0 \to J/\psi K^{*0})
  = [0.85 \pm 0.07(\mathrm{stat}) \pm 0.03(\mathrm{syst})] \times 10^{-3}, \\
\nonumber
\mathcal{B}(B^0_s \to \phi \mu^+ \mu^-)/\mathcal{B}(B^0_s \to J/\psi \phi)
  = [0.90 \pm 0.14(\mathrm{stat}) \pm 0.07(\mathrm{syst})] \times 10^{-3}, \\
\nonumber
\mathcal{B}(B^0 \to K^0_s \mu^+ \mu^-)/\mathcal{B}(B^0 \to J/\psi K^0_s)
  = [0.38 \pm 0.10(\mathrm{stat}) \pm 0.03(\mathrm{syst})] \times 10^{-3}, \\
\nonumber
\mathcal{B}(B^+ \to K^{*+} \mu^+ \mu^-)/\mathcal{B}(B^+ \to J/\psi K^{*+})
  = [0.62 \pm 0.18(\mathrm{stat}) \pm 0.06(\mathrm{syst})] \times 10^{-3}, \\
\nonumber
\mathcal{B}(\Lambda^0_b \to \Lambda \mu^+ \mu^-)/\mathcal{B}(\Lambda^0_b \to J/\psi \Lambda)
  = [2.75 \pm 0.48(\mathrm{stat}) \pm 0.27(\mathrm{syst})] \times 10^{-3}.
\end{eqnarray}

The absolute branching ratios, obtained by substituting the reference 
branching ratios with their PDG~\cite{PDG} values, are
\begin{eqnarray}
\nonumber
\mathcal{B}(B^+ \to K^+ \mu^+ \mu^-)
  = [0.45 \pm 0.03(\mathrm{stat}) \pm 0.02(\mathrm{syst})] \times 10^{-6}, \\
\nonumber
\mathcal{B}(B^0 \to K^{*0} \mu^+ \mu^-)
  = [1.14 \pm 0.09(\mathrm{stat}) \pm 0.06(\mathrm{syst})] \times 10^{-6}, \\
\nonumber
\mathcal{B}(B^0_s \to \phi \mu^+ \mu^-)
  = [1.17 \pm 0.18(\mathrm{stat}) \pm 0.37(\mathrm{syst})] \times 10^{-6}, \\
\nonumber
\mathcal{B}(B^0 \to K^0_s \mu^+ \mu^-)
  = [0.33 \pm 0.08(\mathrm{stat}) \pm 0.03(\mathrm{syst})] \times 10^{-6}, \\
\nonumber
\mathcal{B}(B^+ \to K^{*+} \mu^+ \mu^-)
  = [0.89 \pm 0.25(\mathrm{stat}) \pm 0.09(\mathrm{syst})] \times 10^{-6}, \\
\nonumber
\mathcal{B}(\Lambda^0_b \to \Lambda \mu^+ \mu^-)
  = [1.95 \pm 0.34(\mathrm{stat}) \pm 0.61(\mathrm{syst})] \times 10^{-6}.
\end{eqnarray}

All the numbers are consistent with the B factory measurements~\cite{BFac} 
and enable us to extract NP sensitive quantities from angular observables.

\section{Differential Branching Ratios}
\label{sec:dBR}
\begin{figure}[htb]
\begin{center}
\epsfig{file=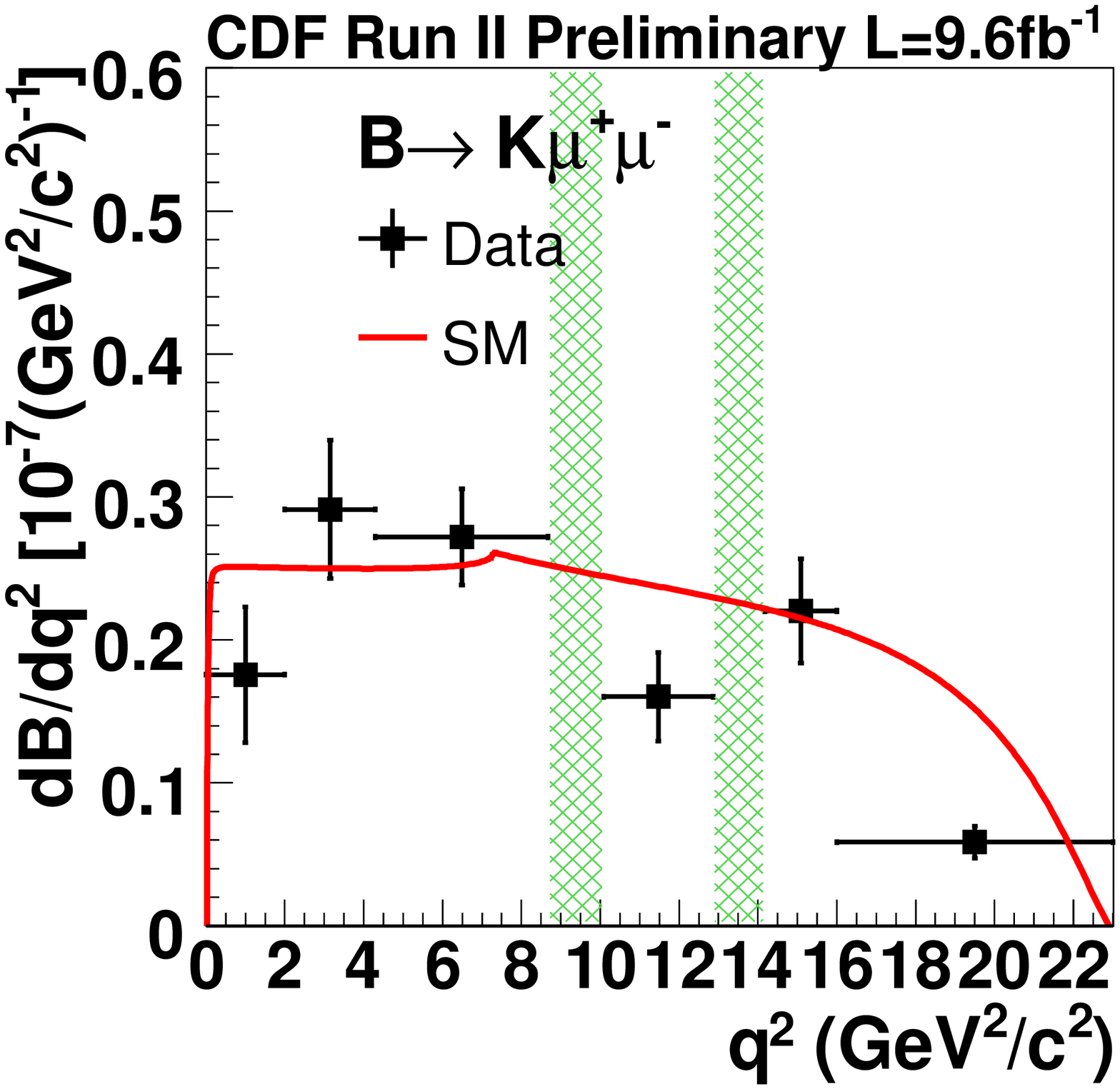,width=0.4\textwidth}
\epsfig{file=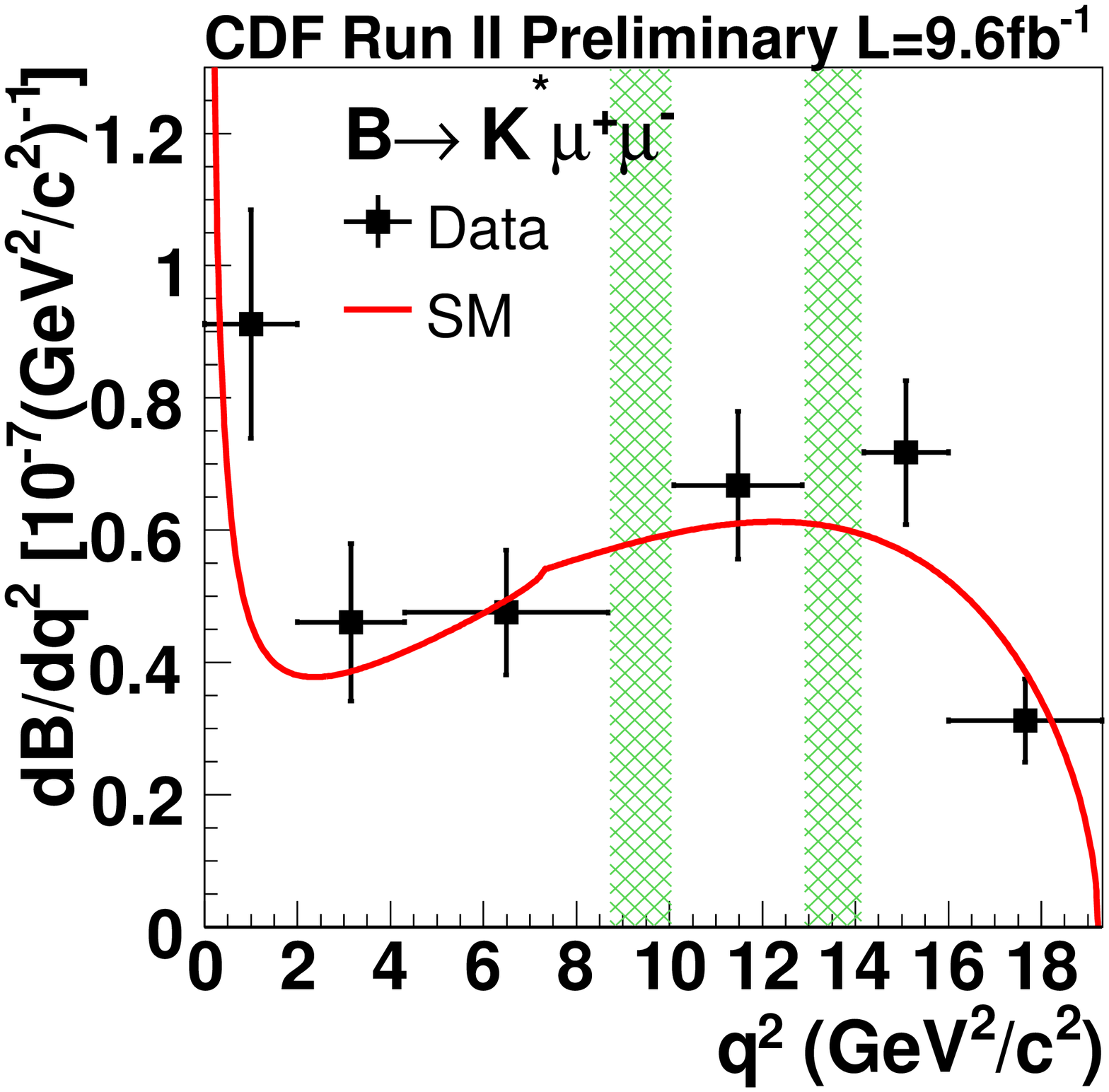,width=0.4\textwidth}
\epsfig{file=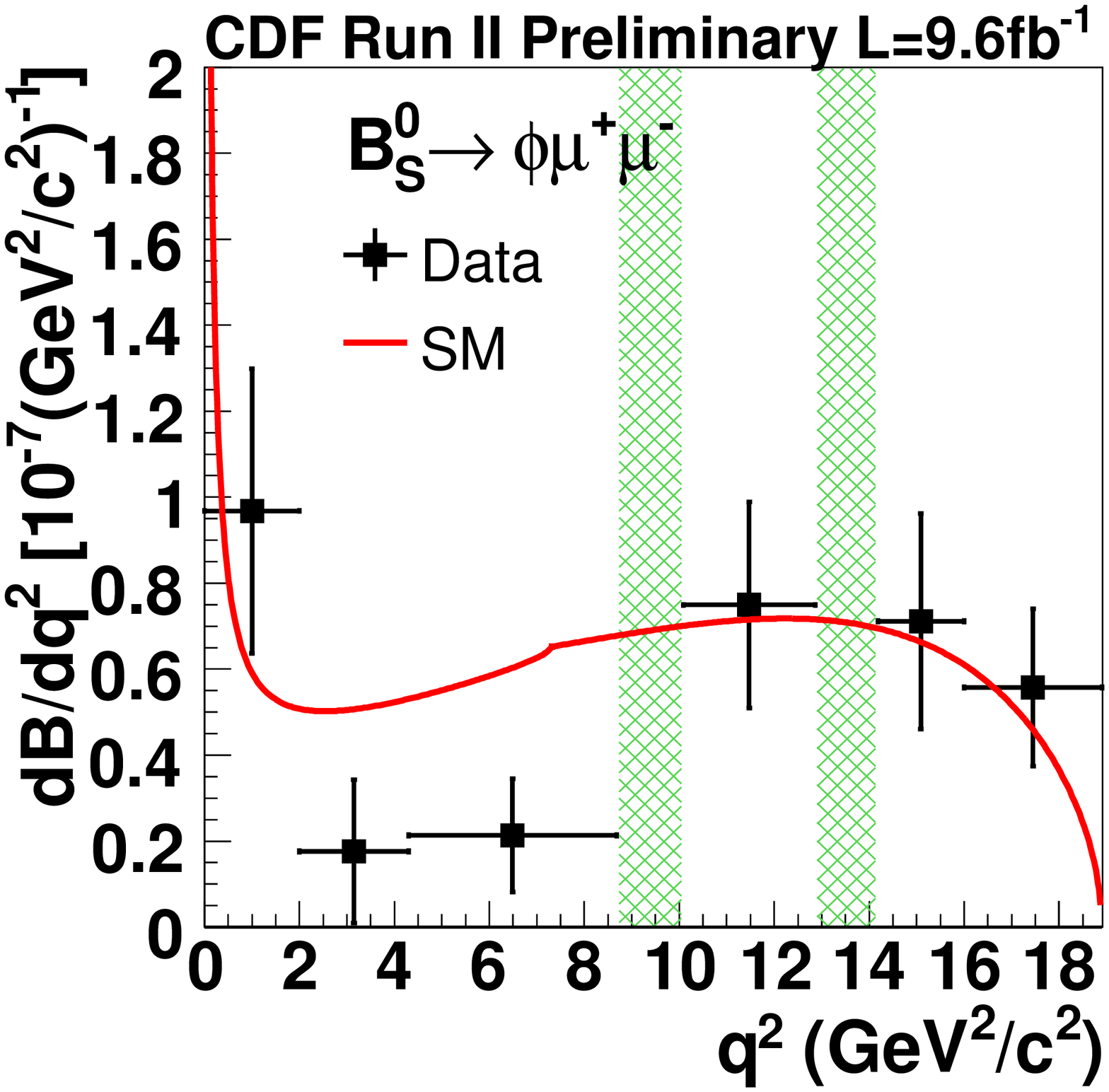,width=0.4\textwidth}
\epsfig{file=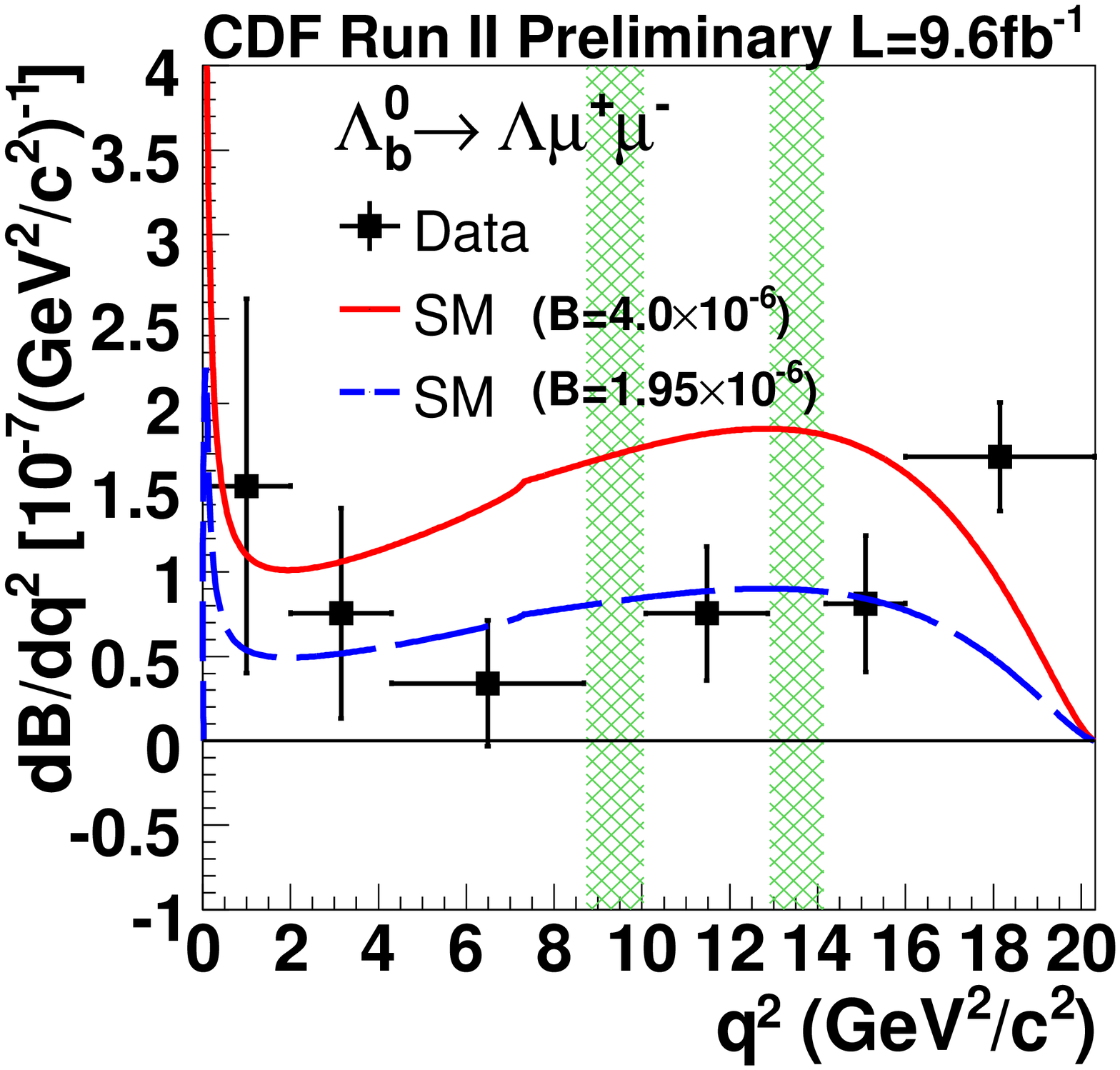,width=0.4\textwidth}
\caption{Differential branching fractions with respect to squared dimuon
mass, $q^2$, in $B \to K \mu^+ \mu^-$ ($K^0_s$ and $K^+$ modes combined),
$B \to K^* \mu^+ \mu^-$ ($K^{*0}$ and $K^{*+}$ modes combined), 
$B^0_s \to \phi \mu^+ \mu^-$, and $\Lambda^0_b \to \Lambda \mu^+ \mu^-$ modes.}
\label{fig:dBR}
\end{center}
\end{figure}

We measure the differential branching ratios with respect to the (squared) 
dimuon mass, $q^2 = M^2_{\mu\mu}c^2$. Same fit procedure as the global fits 
are performed in six exclusive $q^2$ bins to extract the signal yields. In
the fits only the signal fractions are varied, keeping the mean B hadron
masses and the background slopes fixed.
Figure~\ref{fig:dBR} shows the differential branching ratio distributions
for $B \to K \mu^+ \mu^-$ ($K^0_s$ and $K^+$ modes combined), 
$B \to K^* \mu^+ \mu^-$ ($K^{*0}$ and $K^{*+}$ modes combined), 
$B^0_s \to \phi \mu^+ \mu^-$, and $\Lambda^0_b \to \Lambda \mu^+ \mu^-$ modes.
The SM (red curve) predictions are taken from~\cite{DiffBRSM}.
In the $\Lambda^0_b$ plot our data is also compared to the SM prediction 
based on our measured BR value of $1.95 \times 10^{-6}$ (blue dashed curve). 
Also shown, as green vertical bands, are the charmonium veto regions which 
are excluded throughout our analysis. No significant deviations from SM 
prediction are observed.

\begin{figure}[htb]
\begin{center}
\epsfig{file=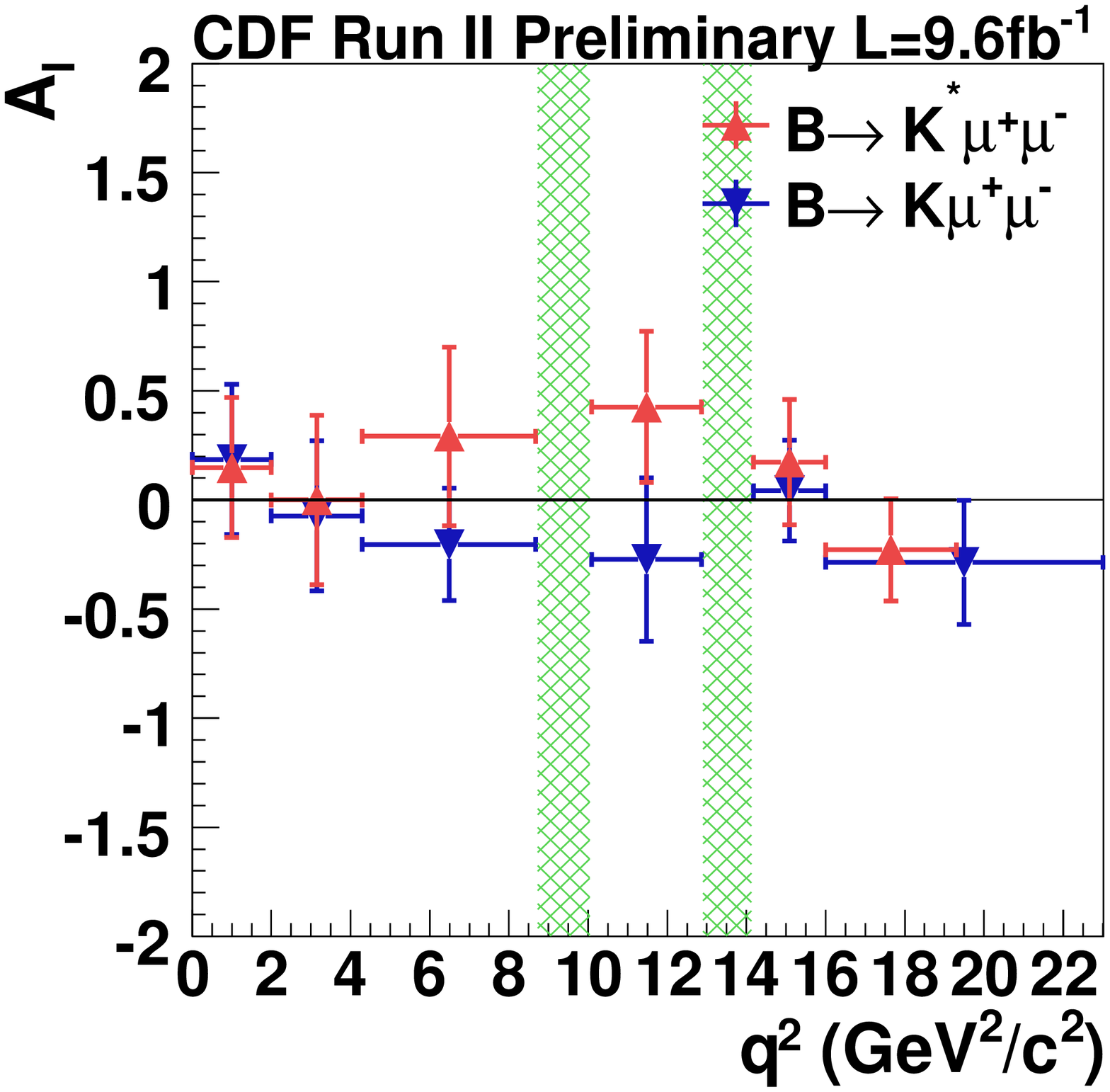,width=0.5\textwidth}
\caption{Isospin asymmetry between neutral and charged B mesons in
$B \to K^* \mu^+ \mu^-$ and $B \to K \mu^+ \mu^-$ modes.}
\label{fig:IsoAcp}
\end{center}
\end{figure}

The isospin asymmetry between the $B^+$ and $B^0$ differential branching 
ratios is defined as, $A_I = [dB(B^0) - r~dB(B^+)]/[dB(B^0) + r~dB(B^+)]$,
where, $1/r$ = $\tau(B^+)/\tau(B^0)$ = 1.071 $\pm$ 0.009~\cite{PDG}, and 
equal production of $B^+$ and $B^0$ is assumed. Figure~\ref{fig:IsoAcp}
shows $A_I$ for the $B \to K \mu^+ \mu^-$ and $B \to K^* \mu^+ \mu^-$ modes.
No significant deviation from zero is observed. We measure the integrated
asymmetries as
\begin{align}
\nonumber
A_I(B \to K \mu^+ \mu^-)
 &= -0.11 \pm 0.13(\mathrm{stat}) \pm 0.05(\mathrm{syst}),\\
\nonumber
A_I(B \to K^* \mu^+ \mu^-)
 &= 0.16 \pm 0.14(\mathrm{stat}) \pm 0.06(\mathrm{syst}).
\end{align}
They are consistent with the B factories and LHCb results~\cite{IsoAcp}.

\section{Angular Analyses of $B \to K^{(*)} \mu^+ \mu^-$ Decays}
\label{sec:Angular}
The differential distributions of the $B \to K^* \mu^+ \mu^-$ 
decays~\cite{AngCorrPheno} are described by four independent kinematic 
variables; the di-muon invariant mass squared ($q^2$),
the angle $\theta_{\mu}$ between the $\mu^+$ ($\mu^-$) direction and the 
direction opposite to the $B$ ($\bar{B}$) meson in the di-muon rest frame, 
the angle $\theta_K$ between the kaon direction and the direction opposite to 
the B meson in the $K^*$ rest frame, and the angle $\phi$ between the two 
planes formed by the di-muon and the $K$-$\pi$ systems. The distributions of 
$\theta_{\mu}$, $\theta_K$, and $\phi$ are projected from the full differential
decay distribution and can be parametrized with four angular observables, 
$A_{FB}$, $F_L$, $A^{(2)}_T$ and $A_{im}$~\cite{AngVars}
\begin{align}
\nonumber
\frac{1}{\Gamma} \frac{d\Gamma}{d\cos\theta_K} & =
  \frac{3}{2} F_L \cos^2 \theta_K + \frac{3}{4} (1-F_L) (1-\cos^2 \theta_K), \\
\nonumber
\frac{1}{\Gamma} \frac{d\Gamma}{d\cos\theta_{\mu}} & =
  \frac{3}{4} F_L(1-\cos^2 \theta_{\mu} + \frac{3}{8} (1-F_L) (1+\cos^2 \theta_{\mu}) + A_{FB} \cos\theta_{\mu}, \\
\nonumber
\frac{1}{\Gamma} \frac{d\Gamma}{d\phi} & =
  \frac{1}{2\pi} [ 1 + \frac{1}{2} (1 - F_L) A^{(2)}_T \cos 2\phi + A_{im} \sin 2\phi ].
\end{align}
where $\Gamma \equiv \Gamma(B \to K^* \mu^+ \mu^-)$, $A_{FB}$ is the muon 
forward-backward asymmetry, $F_L$ is the $K^*$ longitudinal polarization
fraction, $A^{(2)}_T$ is the transverse polarization asymmetry, and $A_{im}$ 
is the triple-product asymmetry of the transverse polarizations.

\begin{figure}[htb]
\begin{center}
\epsfig{file=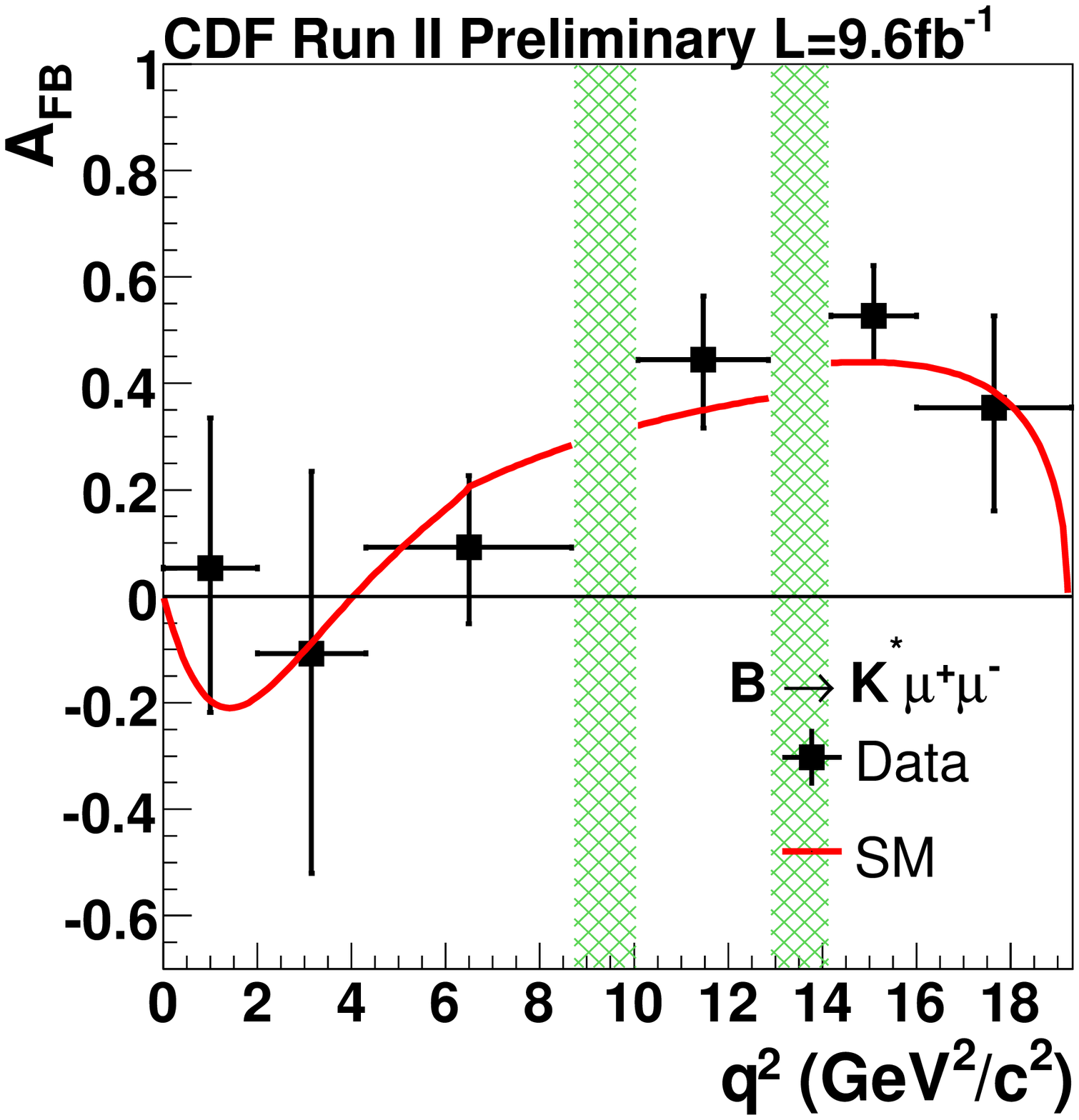,width=0.4\textwidth}
\epsfig{file=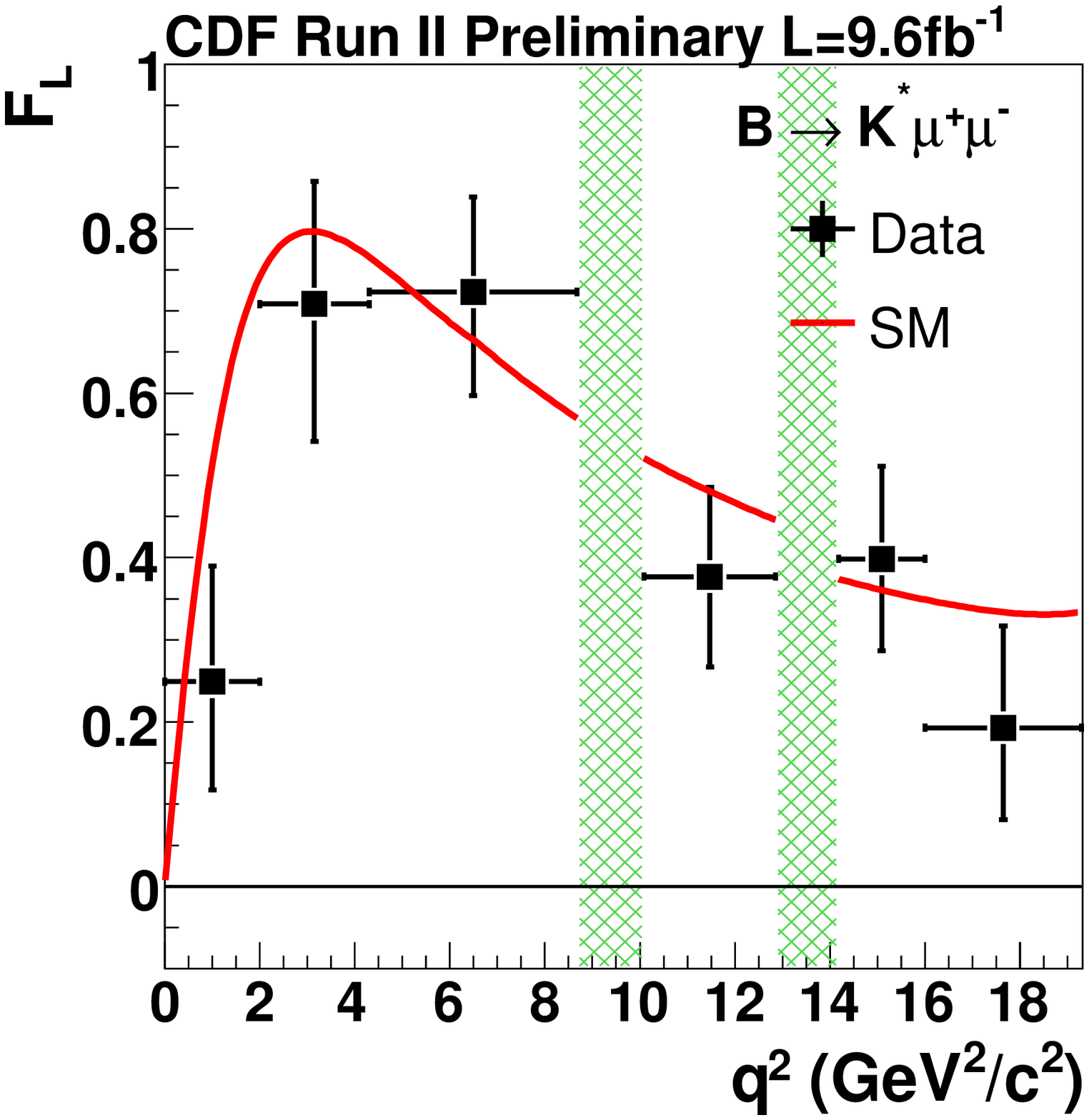,width=0.4\textwidth}
\epsfig{file=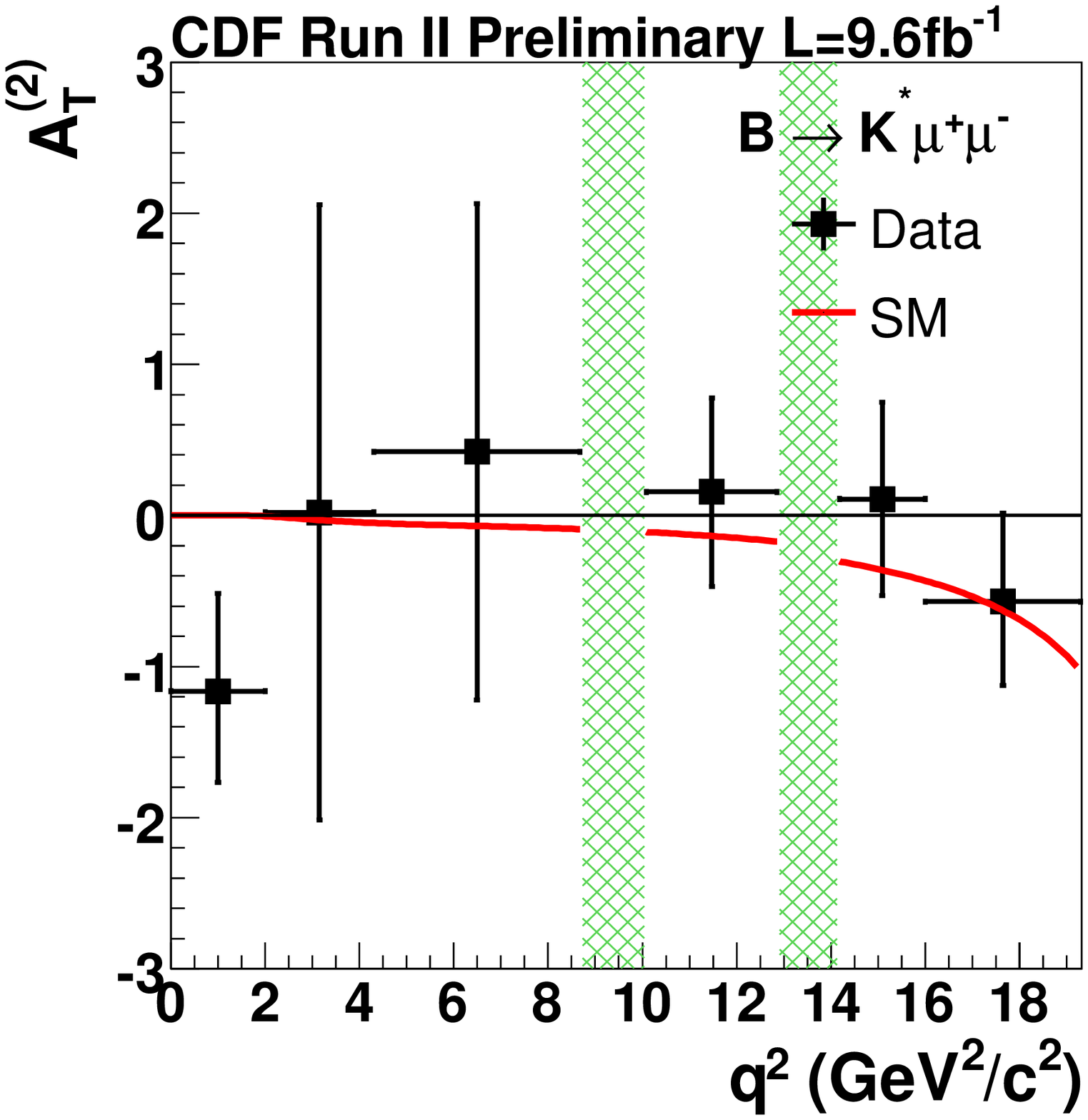,width=0.4\textwidth}
\epsfig{file=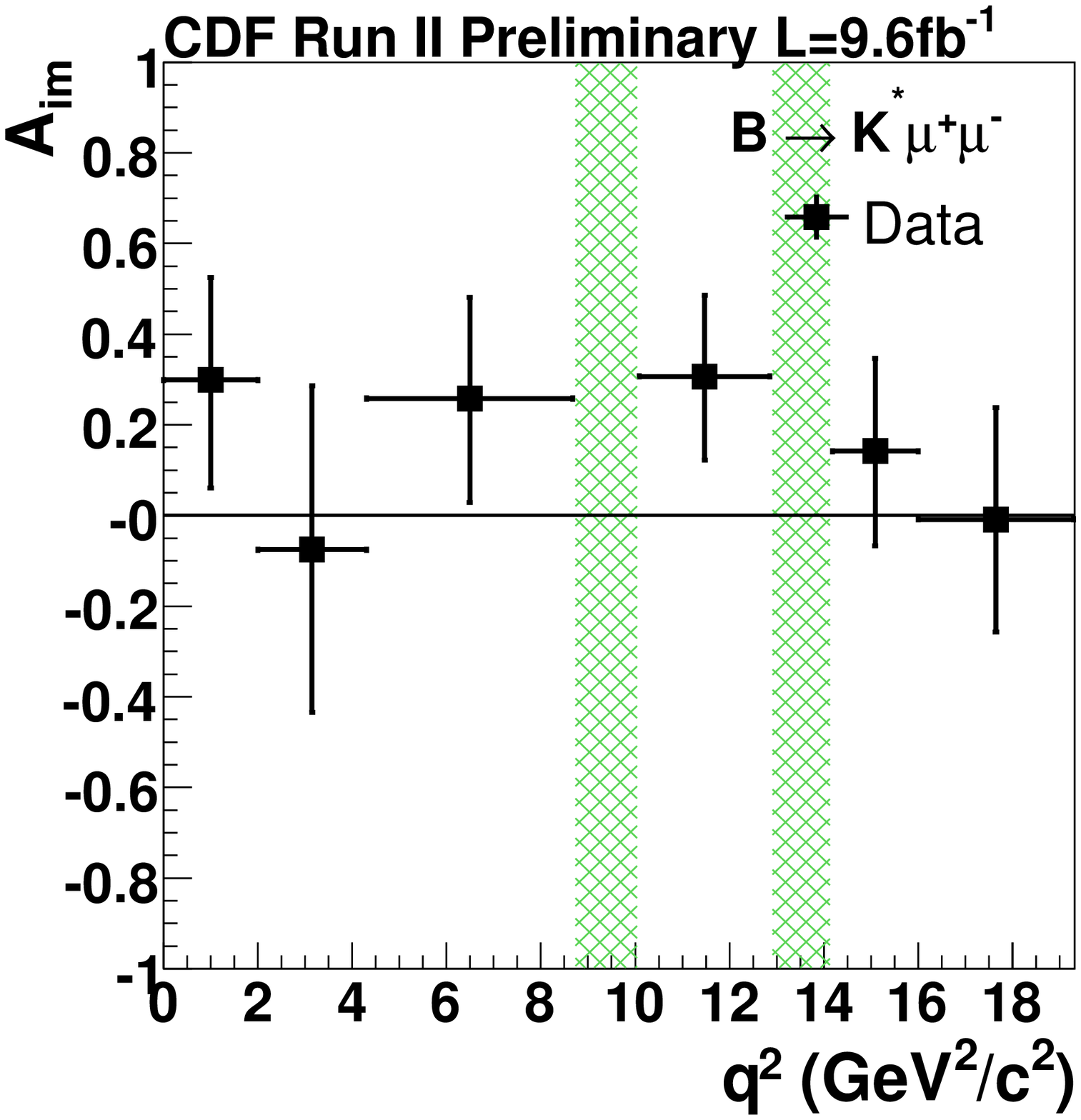,width=0.4\textwidth}
\caption{Angular analysis results of $A_{FB}$, $F_L$, $A^{(2)}_T$ and $A_{im}$ 
with respect to squared dimuon mass, $q^2$, for $B \to K^* \mu^+ \mu^-$ 
decays.}
\label{fig:AngObs}
\end{center}
\end{figure}

We perform an unbinned maximum log-likelihood fit, simultaneously fitting
$K^{*0}$ and $K^{*+}$ in the three angles, $\theta_{\mu}$, $\theta_K$, and 
$\phi$, to extract the four angular observables. Figure~\ref{fig:AngObs} shows
the fitted results with the SM expectations~\cite{EOS}.
We also extract $A_{FB}$ from a similar fit of $B^+ \to K^+ \mu^+ \mu^-$ 
decays, which is consistent with zero as expected.
All the results are consistent with previous measurements and no significant
deviation from SM is observed within current precision.

\section{Summary}
\label{sec:Summary}
We have reported the total and differential branching ratios in 
various $b \to s \mu \mu$ rare decays with the full CDF data sample.
The NP sensitive observables of interest, measured in $B \to K^* \mu^+ \mu^-$
angular analysis, are consistent with standard model expectations and other 
experiments.


\end{document}

%% file: econfmacros.tex
\def\beq{\begin{equation}}
\def\eeq#1{\label{#1}\end{equation}}
\def\eeqn{\end{equation}}

\def\beqa{\begin{eqnarray}}
\def\eeqa#1{\label{#1}\end{eqnarray}}
\def\eeqan{\end{eqnarray}}

\let\bar=\overbar

\def\Dslash{\not{\hbox{\kern-4pt $D$}}}
\def\dslash{\not{\hbox{\kern-2pt $\del$}}}

\def\msb{{\bar{\ssstyle M \kern -1pt S}}}

